\documentclass[superscriptaddress,eqsecnum,amsfonts,showpacs]{revtex4}
\usepackage{epsfig}

\newcommand{\be}{\begin{equation}}
\newcommand{\ee}{\end{equation}}
\newcommand{\bea}{\begin{eqnarray}}
\newcommand{\eea}{\end{eqnarray}}

\newcommand{\nn}{\nonumber}
\newcommand{\ep}{i\epsilon}
\newcommand{\om}{\omega}

\begin{document}

\title{Nakanishi integral representation for  the quark-photon vertex }
\author{V. \v{S}auli}
\affiliation{Department of Theoretical Physics, NPI Rez near Prague, Czech Academy of Sciences}

\begin{abstract}

Using a nonperturbative  framework of Dyson-Schwinger equations a  class  of  Nakanishi's like  integral representations for 
the transverse part of the quark-photon  vertex is  derived.
For this but also for its own purpose the  two and  single variable integral representations for 
 untruncated  quark-antiquark-photon  vertex 
is proposed as well. To exhibit the adequacy of proposed representation,  the Dyson-Schwinger equation
 for the vertex is transformed into the equivalent set of coupled integro-differential equations
for  Nakanishi weight functions-  functions that  appear linearly in the numerator of given integral representation.
Their knowledge then provide self-consistent nonperturbative solution for the vertex in the entire Minkowski space.

  \end{abstract}
\maketitle


\section{Introduction}

The concept of integral representation, which was originally developed for simple scalar theory  by Nakanishi \cite{NAKAN},
 has gradually found its application in more complicated field models. It has  overcome first difficulties  with toy gauge theory, passing the test in  strong coupling QED \cite{saII}  and the first hints appear in practical application in Quantum Chromodynamics, where useful features of Nakanishi's integral
 representation (NIR) were finally appreciated in calculation of meson transition form factors \cite{ding2019}. 

The Dyson-Schwinger equations (DSE) are the equations of motion for Green's functions of the theory and Nakanishi Integral Representation 
is the way to express the Green's functions in very useful manner. 
Using the NIR in   nonperturbative context of Schwinger-Dyson equations is certainly not straightforward task.
Actually, when not looking out of  the Euclidean space,  method based on utilization  of NIR can hardly  compete with  impressive amount of achievements
 based on the straight integration of 
 Dyson-Schwinger equations in the Euclidean momentum space \cite{ROWI194},\cite{MT99},\cite{kacka2005},\cite{WIFIHE2016},\cite{SAWI2018}.
 Such competition  is not why the  NIR is established for and not the main motivation  why  associated  techniques  are developing for.
  Instead of, the  main goal of using  NIR rely on the following facts

 -  since the momentum is very explicit in the expression for NIR,  required space-time transformations, including  Lorentz boost of  vertices and 
meson wave functions turns out to be relatively easy task.
    
-When NIR is used for evaluating  of  form factor, one can  integrate over the loop  momenta analytically.
Consequently, the analytical continuation of calculated form factors can be easily  achieved  and the result can be obtained
 in the entire domain of Minkowski space.  

To appreciate two above points, one obviously has to know NIR.  For this purpose I should point out
the following desired property of NIR

- When  NIR is used properly
in  the tower of DSEs, it  allows the analytical  integration in DSEs and it automatically 
provides analytical continuation of the solution to the entire domain of Minkowski space.

- Regarding the solution of DSEs,  NIR should  be self-consistently self-reproducing.
In other words: when NIR is used to express propagator and vertices inside the DSE for 
n-point vertex, the NIR for this vertex must comes out as a result as well.

The above properties  of NIR were exploited  to provide solutions of many particular 
 quantum field toy model  problems \cite{SA2006}.
Especially, the  NIR was used in  two body bound state calculation in various models 
\cite{KUWI1995},\cite{KUSIWI1997}, \cite{SAAD2013},\cite{SA2008},\cite{KACA2006},\cite{SAFAFR2017},\cite{AC2019}.
 Encouraging results for  the electromagnetic form factors were obtained \cite{CAKA2011} within the formalism as well.

 The first suggestion to use the NIR in QCD is related with  nonperturbative introduction of pinch technique
in pure Yang-Mills theory in 1982 by J. Cornwall \cite{CO1982}.
To the author best knowledge the  functional QCD  resisted against  marriage with the NIR formalism
 for more then 35 years. This problem has been circumvented only recently \cite{VHP} and the
 quark DSE was solved in the  Minkowski space within the use of NIR.
 It provides the  light pion, correct pion decay constant and the  correct width of neutral pion as well as  the first hints  towards the calculation 
of  the off-shell axial anomaly  \cite{PTF}  were made within the formalism.

It is known that transverse pieces of the quark-antiquark-photon vertex could play a crucial role in microscopic QCD explanation
of  Vector Meson  Dominance phenomena. Missing inclusion of  transverse vertices  is apparent in Minkowski space studies \cite{VHP},\cite{PTF}  and it is obvious  drawback of development delay when compared to achievements made in the Euclidean space formalism.
 The main purpose of this paper is to fill this missing gap in our knowledge and  improve the Minkowski space DSEs calculation by gradual inclusion 
 of NIR for transverse  components of the quark-photon vertex. In this paper, within a  newly proposed  NIR, the DSE for the vertex is transformed 
 into the new equation for Nakanishi weight function.  Whilst the  numerical solution of the system is left  for  separate presentation in the near future,
 here it is  the introduction of  NIR, presentation of simple but extendable  QCD DSE  model  and the actual   
 derivation  of NIR for the proper and for the untruncated vertex, which  are main subjects of  three sections of presented paper.

\section{Salam  and Nakanishi  integral representation for  quark-photon vertices}

The original    Perturbation Theory Integral Representation (PTIR) \cite{NAKAN}, when used in nonperturbative context of Dyson-Schwinger equations
must provide the real solution at Euclidean (spacelike) domain of momenta. It comes out form the fact a single  dimensionfull variable which appears in 
the PTIR expression is bounded from bellow.  Also, the so  called Wick rotation is trivially valid \cite{weinberg} and 
 As a consequence, even when  such PTIR is  used in nonperturbative calculations, the solution  in the spacelike domain of momentum
  must be  exactly identical to the one
obtained within the Euclidean theory ( defined by an identical set of  DSE  but  within the Euclidean metric).
 Such property  was many times confirmed in practice (see for instance \cite{saII}),
 however it  leaves a little space to get nonperturbative solution for confining theory like QCD.

It was noticed in \cite{VHP}, that the following  sort of  generalized NIR :
\be 
 \int_{-\infty}^{\infty} d s  \frac{g(s)}{p^2-s+\ep} 
\ee
when used to solve the DSE for  the quark propagator, 
can provide the solution,  which albeit  formally   differs from its 
Euclidean counter-partner solution, can provide physical solution at the end.

For this purpose the  scheme dependent parameters  of truncated  DSEs system
 were  balanced such that the  limit 
\be \label{limit}
 \int_{-\infty}^{\infty} d s  \frac{g(s)}{p^2-s+\ep} 
\rightarrow \int_{0}^{\infty} d s \frac{\sigma(s)}{p^2-s+\ep}
 \, ,
\ee
 was achieved with reasonable numerical accuracy.  
 As one can expect,  gradually vanishing unphysical cut in the quark propagator
  implies the same  for hadron form factors.
  It was exemplified  in \cite{VHP} for the case of   the hadron  vacuum polarization function $\Pi$,
  where it was shown  that analytical properties   assumed long time ago 
   \cite{Cabibbo} can  be restored in the limit (\ref{limit}). 
   In what follow, we allow NIR for  other vertices are unbounded from bellow as well, 
   assuming it can be particularly useful for numerical treatment.

 The integral representation for untruncated vertices can  reduce the number  of integrations necessary  for hadron form  factor evaluations. 
  Hence,  inspired by  algebraic structure  suggested  by Salam \cite{salam63}\cite{stra64}\cite{DEWE1977}\cite{DEWE2}\cite{DE1979} ,
 we propose the integral representation for   complete untruncated quark-photon vertex $G^{\mu}$, which is defined within a complex contour
  $\Gamma$ in the following way 
\bea
 \label{vomba}
&&G^{\mu}(q_-,q_+;Q)=S(q_-)\Gamma^{\mu}(q,Q)S(q_+)
\\
&=&\int_{\Gamma} d x \rho(x) \frac{1}{\not q_- -x}\gamma^{\mu}\frac{1}{\not q_+ -x} 
+\sum_i \int_{\Gamma} d x\int_{-1}^{1} dz  \left[\frac{1}{\not q_- -x}\right]\left[\frac{\rho_i(x) T_i^{\mu}}{q^2+q.Qz+Q^2/4-x^2+\ep}\right]
\left[\frac{1}{\not q_+-x}\right] \, , 
\label{gamba}
\eea
where the reader can recognize the first term in rhs of Eq. (\ref{gamba}), which  is identical to the standard Gauge Technique Anstaz. 
The second term contains  transverse vertices  with the eight momentum dependent matrices $T$, which completes the vertex entirely.  
These matrices  $T$ have one  Lorentz vector index, they satisfy $T.Q=0$,  and they can be chosen to be  Dirac trace orthogonal.
 For a possible choice see for instance \cite{MT99} and we remind here two of them 
\bea
 T_1^{\mu}&=&\gamma^{\mu}-\frac{Q^{\mu}\not Q}{Q^2}
 \nn \\
 T_5^{\mu}&=&q^{\mu}-\frac{q.Q}{Q^{\mu}}{Q^2}
 \nn \\
 \eea 
 which are known to be dominant in Landau gauge 
 ( normalization different from  \cite{MT99} should be used to be able to integrate analytically for general $i$).
Further,   $\rho_i(x)=\rho_i(x;Q^2)$ for $i=1..8$ are corresponding  integral weight functions
, contrary to the function $\rho$, they  depends on
  the square of the photon momentum $Q$, but not on the relative quark momentum $q$.

 The four longitudinal component of the quark-photon vertex function 
 are completely fixed by a single integral  weight function $\rho(x)$ 
 which is identical to the Nakanishi weight in the expression for 
 the quark propagator $S$
 \be \label{spectral}
S(p)=\int_{\Gamma} d x \frac{\rho(x)}{(\not p-x)} \, .
\ee
The  vertex (\ref{gamba}) satisfies the Ward-Takahashi identity:
\be
Q_{\mu}G^{\mu}(q_-,q_+;Q)=S(q_-)-S(q_+) \, ,
\ee
which is a key fact, which allows us to  include Gauge Technique part
into the integral representation  (\ref{gamba}).

 The  proof of the existence of IT (\ref{gamba})  can rely on its explicit finding. 
 We workout Nakanishi Integral Representation  and derive the equations valid for  
 Nakanishi weights for the Ladder-Rainbow truncation of DSEs system. 
  I expect, the integral representation (\ref{gamba}) is valid and should be working for
   more sophisticated approximations as well, if not completely generally.
 
 The solution of the problem thus rely on   converting of the momentum DSE for the vertex into a new  equation for Nakanishi weight functions.
 We will do this in two steps, the first one  is based on the use of IR (\ref{gamba}) and deriving the NIR
 for the proper vertex. In the second step we switched  the obtained  expression  in between two quark propagators 
 and  compare with (\ref{gamba}). In other words,  we write down the NIR for untruncated vertex as well, which   allows us 
 to write down the equations for  functions  $\rho_i$ . Next two sections will be devoted to  two steps just described.

For our quark-photon vertex  DSE we will derive and consider  the NIR for the proper vertex, which reads  
 \bea \label{PTIR0}
 \Gamma^{\mu}(q^+,q_-,Q)&=&\sum_i \int_{-\infty}^{\infty}d a \int_{-1}^{1} dz \frac{O_i^{\mu}\rho_{\Gamma}^i(a,z;Q^2)}
 {F(a,z;q,Q)}
\\  
 F(a,z;q,Q)&=&q^2+zq.Q+Q^2/4-a+\ep \, .
 \eea
where now  capitals $O_i$ state for all 12 matrices needed to describe the vertex in general case. 

Note for completeness,  that the  most general form of NIR based on a very straightforward extension of the  scalar theory PTIR
 is  slightly more complicated and it  would be a little practice here.
 Actually the formula  (\ref{PTIR0}) is completely enough for our purpose, interested reader can find general
 formula  in the Appendix B for any purpose.

Vertex Nakanishi weight $ \rho_{\Gamma}^i(a,z;Q^2)$ is a sum of  distributions in general, 
 which is a fact which must be respected.
  It has a purely continuous part as well as it has pieces,  which contain delta functions. 
 Note, decomposition into selfenergy like terms and the rest  is natural  and it appears already 
in  the  one loop perturbation theory. In order  to make a problem tractable one must decode  the vertex decomposition carefully.
In our case this decomposition  reads 
 \bea \label{PTIR}
 \Gamma^{\mu}(q_-,q_+,Q)&=&\gamma^{\mu}+\sum_i \int_{-\infty}^{\infty}d a \int_{-1}^{1} dz \frac{O_i^{\mu}\rho_{c}^i(a,z;Q^2)}
 {F(a,z;q,Q)}
 \nn \\
 &+&\sum_i \int_{-\infty}^{\infty}d a  \frac{O_i^{\mu}\rho_{\Sigma-}^i(a;Q^2)}{q_-^2-a+\ep}
 +\sum_i \int_{-\infty}^{\infty}d a  \frac{O_i^{\mu}\rho_{\Sigma +}^i(a;Q^2)}{q_+^2-a+\ep}
 +\sum_i \int_{-\infty}^{\infty}d a  \frac{T_i^{\mu}\rho_{\Pi}^{i}(a;Q^2)}{Q^2-a+\ep}\, ,
 \eea
where  now the two variable function  $\rho_{c}(\alpha,\zeta;Q^2)$ as well as  single variable one  $\rho_{\Sigma\pm}(\alpha)$  are
entirely continuous. The tree level $\gamma_{\mu}$ matrix is trivially factorized in the NIR for the proper vertex.

 In QCD, it is plainly believed that the vector mesons factorize as poles in the  BSE solution for the vertex.
In the Eq. we have tentatively added also single variable  
function $\rho_{\Pi}(\alpha)$, which  can be  needed to mimic the vector meson propagators.
 In this paper,  we do not explicitly  factorize $\phi$ or $\omega$ meson  propagator in front of the gauge vertex (\ref{gamba}), 
 if it turns to be necessary,  already presented weight functions are enough for this purpose.  
 With a patience, the resolution should come with planed numerical solution.

\section{ Vertex DSE in the Rainbow-Ladder Approximation }

The DSE for the quark-photon proper  vertex reads
\be \label{vertex}
\Gamma^{\mu}(q,Q)=\gamma_{\mu}+ i\int\frac{d^4k}{(2\pi)^4}G^{\mu}(k,Q)V(k,q) \, ,
\ee
where $G^{\mu}$ is given by the Eq. (\ref{vomba}) and  where  the kernel $V$ is chosen as
\be
V(k,q)_{\alpha\beta\delta\omega}=\gamma^{\mu}_{\alpha\beta}\left[-g_{\mu\nu}V_g(k-q) - g^2\xi \frac{(k-q)^{\mu}(k-q)^{\nu}}{[(k-q)^2]^2}\right]\gamma^{\nu}_{\delta\omega} \, ,
\ee
with the Dirac indices explicit shown. The kernel  is identical to the one  
 in  the  gap equation for the quark propagator: 
\be
S^{-1}(p)=\not p -m-i\int\frac{d^4k}{(2\pi)^4}S(k)V(k,p)
\ee
which has been considered in the paper (\cite{VHP}).

The scalar function $V_g$ has a simple  structure
\be \label{ker}
V_g(k,q)=\frac{c_v(\Lambda^2-\mu^2)}{[(k-q)^2-\mu_g^2][(k-q)^2-\Lambda^2]} \, ,
\ee
which completes our approximation. For completeness, we quote  parameters here
\bea
\alpha_g&=&\frac{c_v}{4\pi}= 22.62,   \frac{g^2\xi}{4\pi}=2.13
\\\nn 
\frac{\mu_g^2}{\Lambda_g^2}&=&0.2667; \mu_g=144 MeV \, ,
\eea
which provide correct  value of  pion mass $m_{\pi}$ and pion decay constant $f_{\pi}$
 via solution of Bethe-Salpeter equation, as well as 
 they provide desired analytical property of hadronic vacuum polarization function. 
 At this place, let us aware  interested reader, that the model here has  also non-confining, 
 but already chiral symmetry breaking phase as was 
 shown in  similar  model studied in \cite{BISA2007}. To show this explicitly,  one needs  non-QCD setup of parameters, e.g. the coupling
should be  few times smaller then quoted here.

Furthermore, within  the use   NIR the extension of (\ref{ker})  is very straightforward.
 Actually, it can  provide an anomalous logarithmic ultraviolet behavior,
as well as   the infrared properties of the kernel can be modified to comply
 with recent DSEs and lattice studies .
 Here, in order to keep the first calculations as  simple as possible,
  we stay  with the form (\ref{ker}) and keep two constant scales 
  $\mu_g$ and $\Lambda_g$ as in the paper (\cite{VHP}).

  In what follows we derive the (\ref{PTIR}) in mean $z$ approximation, where in addition to Gauge Technique part, 
   we will consider the effect of $T_1$ term as the first approximation where transverse vertex is taken into account.
    Explicitly it means that $\rho(x,z)=\rho(x)\delta(z)$ is taken  and thus the following  integral representation 
\bea \label{gamba2}
G^{\mu}(q;Q)&=&\int_{\Gamma} d x    \frac{1}{\not q_- -x}
\left(\gamma^{\mu}\rho(x)+\frac{ \rho_1(x) T_1^{\mu}}{q^2+Q^2/4-x^2+\ep}\right)\frac{1}{\not q_+ -x} \, , 
\eea
is substituted into the vertex  DSE (\ref{vertex}).
 
 To convert DSE into NIR form the integration contour must be specified, which  could be defined at the same domain as the function $\rho$.  
The appropriate contour is represented  by infinite axis cross  in the complex plane of variable $x$, i.e. $\Gamma=\Gamma_1+\Gamma_2$ where
$\Gamma_1:Re x$ , and $\Gamma_2: Im x$ . Making the substitution $o=x^2$ leads to the following  appearance of two common functions
\bea 
g_v(o)&=&\frac{\rho(\sqrt{o})+\rho(-\sqrt{o})}{2\sqrt{o}}
 \nn \\
g_s(o)&=&\frac{\rho(\sqrt{o})-\rho(-\sqrt{o})}{2}
\eea    
defined in the $R^+$ domain of variable $o$, which  gives  plus part of the quark propagator
\be
S_+(p)=\int_0^{\infty} d o \frac{\not p g_v(o)+g_s(o)}{p^2-o+\ep} \, . 
\ee

Considering the contribution from ``unphysical `` contour $\Gamma_2$ one gets
the above functions defined for $o<0$   as superposition of $\rho$ defined  at two branches of square root function of  $x$:
\bea 
g_v(o)&=&i\frac{\rho(\sqrt{-o})-\rho(-\sqrt{-o})}{2\sqrt{-o}}
 \nn \\
g_s(o)&=&-\frac{\rho(\sqrt{-o})+\rho(-\sqrt{-o})}{2} \, ,
\eea  
which gives us the auxiliary function $S_-$ and completes the quark propagator $S=S_- +S_+$ 
\be  \label{dis}
S(p)=\int_{-\infty}^{\infty} do \frac{\not p g_v(o)+g_s(o)}{p^2-o+\ep}  \, .
\ee
In the Eq. (\ref{dis}) we have introduced standard Feynman $\ep$ for usual purpose:  
 the presence of $\ep$ keeps the  analytical continuation  
via Wick rotated contour valid in both cases $S_+$ and $S_-$ and the  integration over the plus  and minus modes  
of $S$ can be performed by  equal footing. 

 Stressed here that  the introduction of integral over negative value modes of $o$ , i.e. the function  $S_-$, should be regarded as an auxiliary step and the function $S-$ is subject of minimization when the system is solved numerically. Absence of $S_-$ would mean the truncation of DSEs system allows to reach an  ideal limit from analyticity point of view.     

Thus analogously for the vertex now, we are going to define two new functions $\tau_{v,s}$ for each component $T_i$ 
\bea 
\tau_v^i(o)&=&\frac{\rho^i(\sqrt{o})+\rho^i(-\sqrt{o})}{2\sqrt{o}}
 \nn \\
\tau_s^i(o)&=&\frac{\rho^i(\sqrt{o})-\rho^i(-\sqrt{o})}{2}
\eea  
for the real positive $o$, and similarly for  negative value of the variable $o$. This two functions are not independent, but  related.

The derivation of  two variable NIR for the proper vertex  is straightforward and 
it follows the receipt described in the Nakanishi's cook book \cite{NAKAN}
The expected subtle  novelty  is that we have to get rid  all terms with relative momentum dependence 
 in the  numerator in a  way they do not show up at the end. This is  feasible task by using a simple algebra combined with per-partes integration and interested reader can find the derivation in Appendix.
Finally,  the NIR for our considered dominant component  $T_1=\gamma_T$ is given by the following expression
\be
\Gamma^{\mu}(q,Q|g,T_1)=V_1^{\mu}(q,Q,m_g)-V_1^{\mu}(q,Q;\Lambda_g)
\ee
where the matrix $V_1$ reads
\bea
V_1^{\mu}(q,Q;m)&=&C \int d \alpha \int_{-1}^{1} d \zeta 
\frac{1}{F(\alpha,\zeta;q,Q)}
\nn \\
&+&\left\{\gamma_T^{\mu} \rho_{\gamma}(\alpha,\zeta)
+ q_T^{\mu} 2\int_0^1 dx (1-x)(1-|\zeta|) \tau_s^{,}(a)\right. \,
\nn \\
&+&\int_0^1 dx 2 q_T^{\mu}
\left(\frac{\not{Q}}{2}(1+\zeta x)+\not q (1-x)\right)(1-|\zeta|)(1-x)\tau^{,}_v(a)
\nn \\
&-& \left.\frac{1}{2}[\not q,\not{Q}]\gamma_T^{\mu}\int_0^1 dx(1-x)(1-|\zeta|)\tau^{,}_v(a)\right\} \, ;
\eea 
where we have used the notation $q_T=T_5$ and  where the weight function matching with  $\gamma^{\mu}_T$ components reads
\bea \label{gam}
\rho_{\gamma}(\alpha,\zeta)&=&\int_0^1 dx \left[-|\zeta|\tau_v(a)
+(1-|\zeta|)\left(a \tau^{,}_v(a)+\tau_v(o)\right) (1-(1-x)^2)\right.
\nn \\
&+&\left.(1-|\zeta|) \tau^{,}_v(a)\frac{Q^2}{4}\left(1+(1-x)^2-2\zeta^2 x+\zeta^2 x^2\right)\right] \, ,
\eea
and   arguments of  $\tau_{v,s}(a)$ ( index$i=1$ was skipped here) is given as 
\be \label{ocka}
a=a(\alpha,\zeta; Q,m;x)=(\alpha-m^2)(1-x)+\frac{Q^2}{4}x(1-\zeta^2) \, .
\ee
Further,  $\tau^{,}(a)$ stands for the  differentiation of the function $\tau$ with respect to the variable $a$. 
The constant $C$ reads  $C=e_q c_v/(4\pi)^2$. 
 
   Most of  terms in the expression above corresponds to weight which  are vanishing at boundaries. There is 
the exception, the first term in   (\ref{gam}), which spoils  this otherwise beautiful property and it gets largest  value 
 for $\zeta=\pm 1$. 

Interestingly also note, that the  function $\tau$ does not rise the single variable dispersion relation, either $\Sigma$ or $\Pi$. 
The reason can be traced back and is due to presence of special   denominator  in  expressions  (\ref{gamba2}) or (\ref{gamba2}).

As there is no such denominator presented  in the case of Gauge Technique IR part of Eq. (\ref{gamba}) ,(\ref{gamba2}) 
the selfenergy like dispersion relation  necessarily arise.
Actually, the Gauge Technique part  of the vertex (\ref{gamba2}) gives rise the following  decomposition of the quark-photon proper vertex NIR 
\bea
\rho_{GT}(\om,\zeta)&=&C \int_0^1 dx  \{[\int^{o_1}_{o_2} d a g_v(a)(1-4x)]
 \\
&-&\frac{g_v(o_1)-g_v(o_2)}{2}(Q^2 x \zeta ((1-x)\zeta+x)\}
 \\
\rho_{GT,\Sigma_+}(\om)&=&C \int_0^1 dx \int^{(\om-\mu_g^2/x)(1-x)}_{(\om-\Lambda_g^2/x)(1-x)} da g_v(a)
\\
\rho_{GT,\Sigma_-}(\om)&=&-C \int_0^1 dx (2x-1) \int^{(\om-\mu_g^2/x)(1-x)}_{(\om-\Lambda_g^2/x)(1-x)} da g_v(a)\, ,
\eea
where $o_1$ and $o+2$ correspond to  the function $o$ (\ref{ocka}) evaluated at the point $m=\mu_g$ and $m=\Lambda_g$ respectively.
We do not show the complete list of results, but only those stemming from the metric tensor in kernel and giving rise 
 (transverse) $\gamma_{\mu}$ matrix structure. 
A more complete  list of the weight functions including also terms stemming  from the the longitudinal term part of  the interaction kernel  
will be published elsewhere.
 
\section{Closing the equations}  

 
 In the previous section two variable NIR for the proper vertex was derived, which is in fact the expression  where 
 the  single variable weight function $\tau$  (and the quark   weights $g$) is contained  in the integral kernel.
 Obviously, how to get the function $\tau$ is not clear yet.
As the last step will rely on some numerical effort, to make the system soluble,
 is equivalent  to writing down  relations where functions $\tau$  can be isolated and singlet out of the integrals.

To do this in practice,  we will use the DSE (\ref{vertex}) sandwiched between quark propagators for this purpose.
 For the left side of the equation  we will use the IR assumed i.e. the Eq. (\ref{gamba})  and convert it into the NIR form but now for $N=3$ in the denominator.

\be \label{tune3}
S(q-)\Gamma^{\mu}S(q^+)=G^{\mu}(q,Q)=\int[d^o_z]
\frac{\sum_i O_i l_i^{[3]}(o,z)}{[F(o,z;k,Q)]^3}
\ee 
and for the rhs of SDE we will do the same
\bea  \label{druha}
&&S(q_-)\left[\gamma_{\mu}+ i\int\frac{d^4k}{(2\pi)^4}G^{\mu}(k,Q)V(k,q)\right]S(q^+)
\nn \\
&=&\int[d^o_z]
\frac{\sum_i O_i r_i^{[3]}(o,z)}{[F(o,z;k,Q)]^3} \, ,
\eea 
where we will use the NIR for the proper vertex (\ref{PTIR}) and in both cases and 
 we will use the IR  (\ref{dis}) to express $S(q_{\pm})$.

Achieving this in practice, sending the left side to the right one  gets the NIR for zero. Since the result is trivial for any $q^2$ and  for all components as well, one simply must have
\be
r_i^{[3]}(o,z)-l_i^{[3]}(o,z)=0 \, .
\ee
which, after a possible  rearrangement give rise  coupled integro-differential equations for  functions $\tau_j$. 
To do this explicitly is a matter of the exercise we present in two following subsection.
For purpose of brevity, starting from this section, we will us  shorthand notations for the NIR measure as already done 
in Eqs. (\ref{tune3}) and (\ref{druha}). Abbreviations are listed  in the  Appendix A .   

\subsection{NIR for the Eq. (\ref{druha})}

We will start  by transforming the second term in lhs of (\ref{druha}) and begin
 with  the $\Pi$-like  dispersion relation (see the Eq. (\ref{PTIR})). For this extraordinary term  
one can  write:

 \bea \label{turn0}
G^{\mu}(k,Q|r,\rho_{\Pi})&=&
\int d o_1 \frac{[(\not k-\not Q/2)g_v(o_1)+g_s(o_1)]}{(k-Q/2)^2-o_1+\ep}
\int d \alpha \frac{ \sum_i O^{\mu i}_{[\Pi]}\rho_{\Pi}^i(\alpha)}{Q^2-\alpha+\ep}
\int d o_2 \frac{[(\not k+\not Q/2)g_v(o_2)+g_s(o_2)]}{(k+Q/2)^2-o_2+\ep}
\nn \\
&=& \int d o_1 d o _2 d \alpha
\int_{-1}^1 dz \frac{\frac{1}{2}[(\not k-\not{Q}/2)g_v(o_1)+g_s(o_1)]
\sum_i O^{\mu i}_{[\Pi]}\rho_{\Pi}^i(\alpha)[(\not k+\not{Q}/2)g_v(o_2)+g_s(o_2)]}
{(k^2+k.Q z+Q^2/4-o_1\frac{1-z}{2}-o_2\frac{1+z}{2})^2(Q^2-\alpha+\ep)} \, ;
\eea

 Let us  match the two denominators in (\ref{turn0}) by the virtue of Feynman identity, such that
\bea
&&(k^2+k.Q z+Q^2/4-o_1\frac{1-z}{2}-o_2\frac{1+z}{2})^{-2} (Q^2/4-\alpha/4+\ep)^{-1}
\\ \nn
&=&\int_0^1\frac{2}{x^2}[k^2+k.Q z+Q^2/4-o_1\frac{1-z}{2}-o_2\frac{1+z}{2}-\alpha\frac{1-x}{4x}+\ep]^{-3}
 \eea
 and perform the  substitution such that 
 $x\rightarrow o$
 \be
 o=o_1\frac{1-z}{2}+o_2\frac{1+z}{2}+\alpha\frac{1-x}{4x} \, .
 \ee
 The result reads
\bea \label{nturn}
G^{\mu}(k,Q|r,\rho_{\Pi})&=&\int[d^o_z]  
\frac{\int do_1 d o_2 d\alpha  N^{\mu}_{\Pi}(\alpha,o_1,o_2)\left[\theta(\alpha)\theta(o-M^2[\vec{o},z])-\theta(-\alpha)\theta(M^2[\vec{o},z]-o)\right]}
{\alpha \left[F(o,z;k,Q)\right]^3}
\eea
where $N^{\mu}_{\Pi}(\alpha,o_1,o_2)$ is identical to  numerator in the integrand in the second line of Eq. (\ref{turn0}) and where
 we have introduced shorthand notation for the  function 
\be \label{mass}
M^2[\vec{o},z]=o_1\frac{1-z}{2}+o_2\frac{1+z}{2}\, .
\ee

Further let us make a sandwich from two variable continuous function $\rho_c(\alpha,\zeta)$  which contributes to  the proper vertex decomposition.
Let us denote the numerator 
 \be
 N^{\mu}_{c}(\alpha,\zeta,o_1,o_2)=[(\not k-\not{Q}/2)g_v(o_1)+g_s(o_1)]\sum_i  T^i \rho^i_c(\alpha,\zeta)[(\not k+\not{Q}/2)g_v(o_2)+g_s(o_2)] 
\ee
and we can immediately write
 \bea \label{turn2}
&&G^{\mu}(k,Q|r,\rho_c)=S(k^-)\Gamma^{\mu}(q,Q|\rho_c) S(k_+)
\nn \\
&=&\int  d \alpha  do_1 do_2 \int_{-1}^{1} d z d \zeta   \frac{\frac{1}{2} N^{\mu}_{c}(\alpha,\zeta,o_1,o_2)  }
{(k^2+k.Q \zeta  +Q^2/4 -\alpha+\ep)(k^2+k.Q z  +Q^2/4 -o_1(1-z)/2-o_2(1+z)/2+\ep)^2}\,  ,
\eea 
 where the first term in the denominator follows from the IR for proper vertex and  we have used the variable $z$ 
 to match  denominators corresponding to quark propagators. 
In what follows, we will  use the Feynman variable $x$ and match two denominators in the expression (\ref{turn2}) together. It gives
us
\be \label{turn3}
\int d \alpha  do_1 do_2 \int_{-1}^{1} d z d \zeta \int_0^1 dx   \frac{ N^{\mu}_{c}}
{(k^2+k.Q (zx+\zeta (1-x))  +Q^2/4 -o_1(1-z)x/2-o_2(1+z)x/2-\alpha(1-x)+\ep)^2}\,  ,
\ee 
then we  continue by  making the substitution $z\rightarrow z^{,}$ such that  $z^{,}=z x+\zeta (1-x)$. 
Further we interchange the ordering of $z^{'}$ and $x$ integration, which gives us
\bea \label{turn4}
G^{\mu}(q,Q|r,\rho_c)&=&\int d \alpha  do_1 do_2 \int_{-1}^{1} d z^{,} d \zeta  
\left[\int_{\frac{z^{,}-\zeta}{1-\zeta}}^1 dx \theta(z^{,}-\zeta) \frac{N^{\mu}_c }{D^3}+\int_{-\frac{z^{,}+\zeta}{1+\zeta}}^1 dx \theta(-z^{,}+\zeta) \frac{N^{\mu}_c }{D^3}\right] \, ;
\nn \\
D&=&k^2+k.Q z^{,}+Q^2/4 -o_1\frac{x(1-\zeta)-z^{,}}{2}-o_2\frac{x(1+\zeta)+z^{,}}{2}-\alpha(1-x)+\ep\,  ,
\eea 
  
Before going further we  simply  relabel $z^{,}\rightarrow z $ and then as the last step
we perform the substitution $x\rightarrow o$ such that 
\be \label{juch}
x=\frac{o-\alpha-o_1 z/2+o_2 z/2}{M^2[\vec{o},z]-\alpha}  \, ,
\ee
where the function $M^2$ is defined above by the rel. (\ref{mass}).
Doing  the substitution explicitly one  gets for the contribution (\ref{turn2})i..e . for the function $ G^{\mu}(k,Q|r,\rho_c)$
the following relation
\be \label{turn5}
\int[d^o_z]   \left[\int d\alpha  do_1 do_2   \int_{-1}^{1}  d \zeta  N^{\mu}_c(\alpha,\zeta,o_1,o_2)
\frac{\theta(z-\zeta)\theta(1-x)\theta(x-\frac{z-\zeta}{1-\zeta})+\theta(\zeta-z)\theta(1-x)\theta(x-\frac{\zeta-z}{1+\zeta})}
{\left[M^2[\vec{o},\zeta]-\alpha\right][F(o,z;k,Q)]^3}\right] \, .
\ee
In words: contributions to  $r_i$ can be readily identified from derived relation above and they are given by the  four dimensional integral
over the three Nakanishi weights in this special case.
 

Without large effort one can write down last  expressions for the part of semi-amputated  $G^{\mu}(k,Q;r,\Sigma_{\pm})$ , 
which arises in the rhs. of vertex DSE and is due to the single variable $\rho_{\Sigma_{\pm}}$ (hence the labeling).
The expressions follow as limit $\zeta=\pm 1$ of the above relation (\ref{turn5}). They are following: 
\bea \label{turn6}
G^{\mu}(k,Q|r,\Sigma_+)&=&\int[d^o_z]    \frac{\int d\alpha  do_1 do_2  }{[F(o,z;k,Q)]^3}
\frac{\theta(1-x_+)\theta(x_+ -\frac{1-z}{2})N^{\mu}_{\Sigma_+}(\alpha,o_1,o_2)}{o_2-\alpha}
\eea
\bea \label{turn7}
G^{\mu}(k,Q|r,\Sigma_-)&=&\int[d^o_z]    \frac{\int d\alpha  do_1 do_2  }{[F(o,z;k,Q)]^3}
\frac{\theta(1-x_{-})\theta(x_{-} -\frac{1+z}{2})N^{\mu}_{\Sigma_-}(\alpha,o_1,o_2)}
{o_1-\alpha}
\eea 
and where functions $x_{\pm}$ are given by (\ref{juch}) evaluated at $\zeta$ endpoints,  explicitly written 
\bea \label{juch2}
x_{+}=\frac{o-\alpha-o_1 z/2+o_2 z/2}{o_{2}-\alpha}  \, ,
\nn \\ 
x_{-}=\frac{o-\alpha-o_1 z/2+o_2 z/2}{o_{1}-\alpha}  \, .
\eea
 
 The linear  presence of the weight function $\rho_{\Sigma}(\alpha)$ in the matrices $N^{\mu}_{\Sigma}$
 is
 \be
 N^{\mu}_{\Sigma}(\alpha,o_1,o_2)=[(\not k-\not{Q}/2)g_v(o_1)+g_s(o_1)]\sum_i O^{\mu}_{i}\rho_{\Sigma}^i(\alpha)[(\not k+\not{Q}/2)g_v(o_2)+g_s(o_2)] \, .
\ee

In the approximation we employed, the rhs of DSE for the vertex is converted
to the desired integral expression (\ref{druha}) given by the sum
\be
G^{\mu}(q,Q|r,\Sigma_-)+G^{\mu}(q,Q|r,\Sigma_+)+G^{\mu}(q,Q|r,\rho_c)+G^{\mu}(k,Q|r,inh.) \, ,
\ee
where the last term corresponds to  the inhomogeneous term, i.e. the gamma matrix sandwiched in between propagators
This term reads
\bea \label{N3}
 G^{\mu}(k,Q|r,inh)&\equiv&S(k_-)\gamma^{\mu} S(k_+)
=\int[d^o_z] \frac{\int do_1 do_2 N^{\mu}_{inh}(o_1,o_2) \theta(o-M^2(\vec{o},z))}{F^3(o,z;k,Q)}
  \\
N^{\mu}_{inh}(o_1,o_2)&=&[(\not k-\not{Q}/2)g_v(o_1)+g_s(o_1)]\gamma^{\mu}[(\not k+\not{Q}/2)g_v(o_2)+g_s(o_2)]
\eea
with the short derivation  delegated into the Appendix.

\subsection{$N=3$ NIR for the Eq. (\ref{tune3})} 

Rewriting the l.h.s. of means nothing else but showing that our two Anstaz \ref{gamba2} (or more generally (\ref{gamba}))
complies with the NIR as well. As we are dealing with untruncated vertex, which is the equivalent of the Bethe-Salpeter wave function
we have chosen $N=3$ for this purpose.  
  
 Therefore  we rewrite our assumed integral representation for un-amputated vertex into this form.
 
 To achieve $N=3$ NIR for the Gauge Technique Anstaz  one needs to match the propagator together and perform one 
 per-partes integration, which gives us
 \be 
G^{\mu}(q,Q|l,GT)=- \int[d^o_z]   \frac{ g_v^{[0]}(o)[(\not k-\not Q/2)\gamma^{\mu} 
(\not k+\not Q/2)+\gamma^{\mu} g_v^{[1]}(o)]+\left(2 k^{\mu}+\frac{1}{2}[\gamma^{\mu},\not Q]\right) g_s^{[0]}(o)}
{[F(o,z;k,Q)]^3}\,  ,
\ee 
 where we have labeled zero (primitive function) and the first  momentum integral of the function $g$ as
 \bea \label{pica}
 g^{[0]}(o)&=&\int_{-\infty}^o g(x) dx \, 
\nn \\ 
 g^{[1]}(o)&=&\int_{-\infty}^o x g(x) dx \, .
 \eea

The NIR for transverse parts can be cast into the form
\be \label{turning1}
G^{\mu}(q,Q|l,T)=- \int[d^o_z] \sum_i
\frac{N^{\mu,i}(o)(1-|z|)}{[F(o,z;k,Q)]^3} \, ,
\ee
where the matrices in the numerator reads
\be
N^{\mu, i}(o)= \tau_v^i(o)
 \left[(\not k-\not{Q}/2)T_i^{\mu}(\not k+\not{Q}/2)+ o T_i^{\mu}\right]+ 
\tau_s^i (o)\left[(\not k-\not{Q}/2)T_i^{\mu}+T_i^{\mu}(\not k+\not{Q} /2)\right] \, .
\ee 

The derivation consists from matching of three denominators that appear in  the IR (\ref{gamba2}) and is equivalent to  performance that has led
to the expression (\ref{mezihra})  in the Appendix.

   \section{Conclusion}

 The inhomogeneous Bethe-Salpeter equation for the quark photon-vertex has been transformed into the equivalent integro-differential equations for the Nakanishi weight functions. The Nakanishi representation has been found for the vertex, which in addition to the part given by the  Gauge Technique, involves
 the dominant transverse component $\gamma_T$ in a newly proposed integral representation for untruncated vertex. 

  Either a novel representation for non-amputated vertex, as well as  the derived Nakanishi representation for the proper vertex show that
 there are cuts in multivariable complex hyperplane associated with various momentum dependence of the vertex, e.g. with momentum carrying  by 
 quarks  represented by external legs. The presence of cuts lie behind the fact why the Dyson-Schwinger equations or Bethe-Salpeter equations are not numerically solvable in momentum Minkowski space. Transformed equation within the DSE do not share this property 
 and their numerical solutions are well defined. 
Albeit  transforming the DSEs into the new equations for Nakanishi weight  can represent not easy and sometimes quite  demanding job, the practical 
future use in calculations of  hadron production amplitudes in Minkowski space is worthwhile to following this path.

\appendix

\section{\\ Shorthand notations and often used conventions}

The author uses Minkowski metric $g_{\mu\nu}=diag(1,-1,-1,-1)$ hence 
$[\gamma^{\mu},\gamma^{\nu}]=2g^{\mu\nu}$.

Multidimensional integration if repeated with the same boundaries we write

 $$\int d \alpha f=:  \int_{-\infty}^{\infty} d \alpha f $$
  
$$  \int d \alpha_1 d \alpha_2...d \alpha_n f(\vec{\alpha},..) =:  \int_{-\infty}^{\infty} d \alpha_1  \int_{-\infty}^{\infty} d \alpha_2...  \int_{-\infty}^{\infty} d \alpha_n f(\vec{\alpha},..)$$ .

The following dense notation for the integration over the infinite strip
$$\int[d^a_z]  f =: \int_{-\infty}^{\infty} d a  \int_{-1}^{1}  d z  f $$  
is used for brevity, where $f$ is an arbitrary function to be integrated over.

\section{NIR for the vertex}
 
The most general 3-variable Nakanishi integral representation, which for a scalar vertex reads
 \be \label{TIR}
 \Gamma_{S}(p,p',q)=\int_{0}^{\infty}d a \prod_{i=1..3} \int_{-1}^{1} dz_i \frac{\delta(1-\sum_{i=1}^3 z_i)  \rho(a,\vec{z})}
 {z_1p^2+z_2p^{'2}+z_3q^2 -a+\ep }\, ,
 \ee
since $\rho(a,\vec{z})$ is a sum of product of  various distributions, obviously  the object not much  useful in practical calculation.
For QED proper vertex there  is a twelve such integrals, the four are  associated with longitudinal components and
 the eight with the transverse components.

\section{ Derivation of the NIR for  $\Gamma^{\mu}(k,Q|T_1,g)$}
\label{hrdybud}

In this Appendix we derive the $N=1$  NIR for the contribution stemming from the metric
 tensor part of the interaction kernel and due to  the $T_1=\gamma_T$ 
 component of the transverse part of the vertex. 
 
In fact it is enough to consider only  $\gamma_{\mu}$ matrix, since the complete result can  obtained by 
contracting the result with the tensor $P_T^{\mu\nu}=g_{\mu\nu}-\frac{Q^{\mu\mu}}{Q^2}$.
To establish convention,  for the considered contribution we can write
\be \label{subtr}
\Gamma(q,Q|T_1,g)^{\mu}=P_{T\nu}^{\mu}\Gamma_1^{\nu}(q,Q) \,
\ee
where thus we only  need to consider the following term 
\bea \label{dot}
\Gamma_1^{\mu}(q,Q)&=&I(q,Q;\mu_g)-I(q,Q;\Lambda_g)=
\nn \\
I(q,Q;\mu_g)&=&i\int\frac{d^4k}{(2\pi)^4} \int da \frac{2\gamma^{\mu}(k^2-\frac{Q^2}{4})-2\gamma^{\mu} a
- 4(k^{\mu}-\frac{Q^{\mu}}{2})(\not k-\frac{\not{Q}}{2})-[\not k,\not{Q}]\gamma^{\mu}\tau_v(a)+
+ 4(k^{\mu}+\frac{Q^{\mu}}{2})\tau_s(a)}{D_1D_2D_3D_4} \, \;
\\
D_1&=&\Delta(k-Q/2;a)\, \, \, ;\,\,\, D_2=\Delta(k+Q/2;a)\, \, \, ;\,\,\, D_4=\Delta(k-q;m_g^2) \, \, \, ;\,\,\, D_5=\Delta(k-q;\Lambda_g^2) 
\nn \\
D_3&=&k^2+Q^2/4-a+\ep \, ,
\eea 
where  within the use of the identity $ [\gamma_{\mu},\gamma_{\nu}]=2g_{\mu\nu}$  a little algebraic   rearrangement was 
made in the numerator of the Eq. (\ref{dot}) and where Greek $\Delta$ stands for the scalar Feynman propagator in the momentum space: 
$\Delta(k,o)=k^2-o+\ep; \, \epsilon >0$ .

In the beginning,  it is useful to rewrite the IR for the vertex inside the loop into the $N=3$ NIR . Doing this explicitly it allows to get rid of one auxiliary  Feynman integral even before the momentum integration comes in order. It requires to match  all $D$'s,   which involve the variable $a$. A possible sequence of steps reads 
 \be
 \frac{1}{D_1 D_2}=\frac{1}{2}\int_{-1}^{1} dy \frac{1}{[k^2+k.Q y+\frac{Q^2}{4}-a+\ep]^2}
 \ee
 \be
 \frac{1}{D_1 D_2 D_4}=\int_{-1}^{1}  dy\int_0^1 dx  \frac{x}{[k^2+k.Q y x+\frac{Q^2}{4}-a+\ep]^3}
 \ee
 At this point we make a substitution $y\rightarrow z$ such that $z=yx$ and interchange the integration ordering. This gives the following 
 double integral
 \be
 \frac{1}{D_1 D_2 D_4}=\int_{-1}^{1}  d z\frac{\int_z^1 dx \theta(z)+\int_{-z}^1 dx \theta(-z)}{[k^2+k.Q z+\frac{Q^2}{4}-a+\ep]^3} \, ,
 \ee
 where the integration over the variable $x$ can be easily done, showing thus that  just a  single variable is enough to match all
  denominators  of the IR for the untruncated transverse vertices. This important  result reads
  \be \label{mezihra}
 \frac{1}{D_1 D_2 D_4}=\int_{-1}^{1}  dz  \frac{1-|z|}{k^2+k.Q z+\frac{Q^2}{4}-a+\ep}^3 \, .
 \ee

 Here , we will use it further and we match the result (\ref{mezihra}) with   $1/D_3$ by using a new Feynman variable $x$ for this purpose.
 The procedure is straightforward: we  complete the square in the integration momentum, perform  a standard shift and integrate over the  momentum.
  For the result, again back in Minkowski  momentum space, we can write
\bea \label{huba}
\Gamma_1^{\mu}(q,Q)&=&i\int da \int_{-1}^{1} dz \int_0^1 dx \frac{-x^2(1-|z|)}{(4\pi)^2}
\left(\gamma^{\mu}\frac{\tau_v(a)}{D}+\frac{N^{\mu}}{D^2}\right)-[\mu_g\rightarrow \Lambda_g]
\nn \\
N^{\mu}&=&\gamma^{\mu}\tau_v(a)\left[(-\frac{Q}{2} z x+q(1-x))^2-(a+\frac{Q^2}{4})\right]
\nn \\
&-&2\tau_v(a)\left[-\frac{Q^{\mu}}{2} (z x+1)-q^{\mu}(1-x)][\frac{\not{Q}}{2}(zx+1)+\not q (1-x)\right]
\nn \\
&-&\frac{\tau_v(a)}{2}[\not q,\not Q]\gamma^{\mu}(1-x)
\nn \\
&+&2\tau_s(a)\left[-\frac{Q^{\mu}}{2} (1-z x)+q^{\mu}(1-x)\right]
 \\
D&=&x(1-x)q^2+q.Qz x (1-x)+\frac{Q^2}{4}(1-z^2) x- a x-\mu_g^2(1-x)
\eea

Factorizing $ x(1-x)$ from the denominator $D$ and making the substitution $a \rightarrow o$
such that
\be
-o=-\frac{Q^2}{4}+\frac{Q^2}{4}\frac{1-z^2 x}{1-x}- \frac{a}{1-x}-\frac{\mu_g^2}{x}
\ee
one arrives at the following expression for the relation (\ref{huba}) 

\bea 
\Gamma_1(q,Q)&=&i\int do \int_{-1}^{1} dz \int_0^1 dx \frac{-x (1-|z|)}{(4\pi)^2}
\left(\gamma^{\mu}\frac{\tau_v(a)}{F(o,z;q,Q)}\right.
\nn \\
&+&\left.\gamma^{\mu}\frac{\tau_v(a)}{x(1-x)[F(o,z;q,Q)]^2}
\left[(q^2(1-x)^2-q.Qz x (1-x) +\frac{Q^2}{4} z^2 x^2-a-Q^2/4)\right]\right)
+... -[\mu_g\rightarrow \Lambda_g]
\nn \\
a&=&(o-\frac{\mu_g^2}{x})(1-x)+\frac{Q^2}{4}x(1-z^2) \, ,
\label{acko}
\eea
 where three dots represent last three lines of $N^{\mu}$ in the relation (\ref{huba})  divided by the function $x(1-x)^2 [F(o,z;q,Q)]^2$.
 
In what follow, we use the algebraic identity  for  the term $q^2/F^2$ 
 \be  
 \frac{q^2}{F(o,z,q,Q)^2}=\frac{1}{F(o,z,q,Q)}-\frac{q.Q z+\frac{Q^2}{4}-o}{[F(o,z;q,Q)]^2}
 \ee  
 and for all terms proportional to $1/F^2$,  with the  exception of those involving variable  $q.Q$,
  we apply  per-partes integration in the variable $o$ and reduce  thus negative  power of $F$ from two  to one.
   
 Then the only  reminding  terms that need to be  transformed into $N=1$ form of desired NIR, are those  proportional to $q.Q/F^2 $.
We collect them and use  the per-partes integration with respect to the variable $z$.
 More explicitly, the appropriate per-partes has vanishing boundary term and explicitly reads
 \be
\int_{-1}^{1}dz \frac{(1-|z|) z q.Q}{[F(o,z;q,Q)]^2}\tau_v(a)=
\int_{-1}^{1}dz\left[\frac{(1-2|z|)}{F(o,z;q,Q)}\tau_v(a)+
 \frac{(1-|z|) }{F(o,z;q,Q)}\frac{d\tau_v(a)}{da}\frac{da}{dz}\right]
 \ee
 where $da/dz=-\frac{Q^2}{2}x z$.
 
 Putting all together one gets the final result:
 \bea
\Gamma_1^{\mu}(q,Q)&=&\int d o \int_{-1}^{1} d z \frac{1}{F(o,z;q,Q)}
\left[\gamma^{\mu} \rho_{\gamma}(o,z)\right.
\nn \\
&+&\left(\frac{\not{Q}}{2}(1+z x)+\not q (1-x)\right)\frac{d\tau_v(a)}{d a}
\nn \\
&-&\int_0^1 dx \frac{1}{2}[\not q,\not{Q}]\gamma^{\mu}(1-x)\frac{d \tau_v(a)}{d a}
\nn \\
&+&2\left.\int_0^1 dx \left(\frac{Q^{\mu}}{2}(1-z x)+q^{\mu}(1-x)\right)\frac{d\tau_s(a)}{d a}\right] \ ,  ,
\eea 
with the weight function corresponding to the  $\gamma^{\mu}$ matrix singlet out for better recognition 
\bea
\rho_{\gamma}(o,z)&=&\int_0^1 dx \left[-|z|\tau_v(a)
+(1-|z|)\left(a \frac{d \tau_v(a)}{d a}+\tau_v(a)\right) (1-(1-x)^2)\right.
\nn \\
&+&(1-|z|) \frac{d \tau_v(a)}{da}\frac{Q^2}{4}\left(1+(1-x)^2-2z^2 x+z^2 x^2\right)] \, ,
\eea
with the argument of functions  $\tau_{v,s}(a)$  defined in the Eq. (\ref{acko}). 

The complete result obtained by subtraction (\ref{subtr}) is reviewed in the main text.

\section{$N=3$ NIR for the product $S(k^+)\gamma^{\mu} S(k^-)$}

In this appendix the product of two propagators is transformed into NIR, which has  the third power of $F$ in the denominator.
As the first step, let us perform  per partes integration for one of two quark propagators. 
Choosing the left one  for this purpose one gets
 \bea \label{N31}
S(k_-)\gamma^{\mu} S(k_+)
=\int dx_1 dx_2 \frac{[(\not k-\not{Q}/2)g_v^{[0]}(x_1)+g_s^{[0]}(x_1)]\gamma^{\mu}[(\not k+\not{Q}/2)g_v(x_2)+g_s(x_2)]}
{[(k_-)^2-x_1+\ep]^2[(k_+)^2-x_2+\ep]} \, ,
\eea
where we have used the notation (\ref{pica}) for convenient realization  of  primitive functions.
 We assume the boundary term vanishes and the limit
\be
\lim_{o\rightarrow \infty}\frac{g_{v,s}(o)^{[0]}}{o}=0
\ee 
is satisfied, noting the numerator in the limit  should be a finite constant and the limit is obviously achieved.
Matching denominators  together by using the variable $z$, one gets for (\ref{N31}) following expression
\be \label{N32}
\int_{-1}^{1} dz\int dx_1 dx_2 (1-z)
\frac{[(\not k-\not{Q}/2)g_v^{[0]}(x_1)+g_s^{[0]}(x_1)]\gamma^{\mu}[(\not k+\not{Q}/2)g_v(x_2)+g_s(x_2)]}
 {[k^2+k.Qz +Q^/4-x_1(1-z)/2-x_2(1+z)/2+\ep]^3}
\ee

Making the substitution $x_1\rightarrow o$ such that
\be
o=x_1(1-z)/2+x_2(1+z)/2
\ee
one gets for (\ref{N32})
\be \label{N33}
\int_{-1}^{1} dz\int do dx_2 \int_{-\infty}^{o_1^{max}} d o_1 \frac{[(\not k-\not{Q}/2)g_v(o_1)+g_s(o_1)]\gamma^{\mu}[(\not k+\not{Q}/2)g_v(x_2)+g_s(x_2)]}
{[k^2+k.Qz +Q^2/4-o+\ep]^3}
\ee
where the upper boundary of the third integral reads
\be
o_1^{max}=x_1=\frac{2}{1-z}\left(o-x_2\frac{1+z}{2}\right)
\ee
Now, by a simple relabeling $x_2\rightarrow o_2$ and expressing the boundary in a form of Heaviside theta function
one immediately arrives into the desired expression (\ref{N3}).

The identity (\ref{N3}) was derived by using the expression (\ref{pica}), which is  obviously derived for continuous functions. 
Not surprisingly, it is valid for the perturbation theory distribution as well. This fact  can be inspected by taking 
$g_v(x)=\delta(x-m^2)$ and $g_s(x)=m\delta(x-m^2)$ 
 in the expression (\ref{N3}) and by performing  the integration over the variables $o$ and $z$. As it should be, we recover
 the expected product of free fermion propagators with  $\gamma^{\mu}$ matrix in between.


%

\begin{thebibliography}{00}

\bibitem{NAKAN}
Nakanishi N. , Graph Theory and Feynman Integral (Gordon and Breach, New York, 1971).

\bibitem{saII}
V. Sauli, JHEP 0302:001,2003 .

\bibitem{ding2019}
M. Ding, et. al.; Phys. Rev. D99,014014 (2019).

 \bibitem{ROWI194}
C. D. Roberts,  A. G. Williams, Prog. Part. Nucl. Phys. 33, 477-575 (1994).

\bibitem{MT99}
P. Maris, P.C. Tandy, Phys. Rev. C61, 045202 (2000).

\bibitem{kacka2005}
P. Maris,  P.C. Tandy, Phys. Rev. C62, 055204 (2000). 
 
 \bibitem{WIFIHE2016}
R. Williams, C. S. Fischer, W. Heupel, Phys. Rev. D 93, 034026 (2016).
  
 
\bibitem{SAWI2018}
Hèlios Sanchis-Alepuza, R. Williams, Computer Physics Communications 232, 1-21, (2018).
 
 \bibitem{SA2006}
V. Sauli, Few Body Syst. 39, 45 (2006).  
    
\bibitem{KUWI1995}
K.  Kusaka, A. G.  Williams, Phys. Rev. D51, 7026 (1995).
 
\bibitem{KUSIWI1997}
K. Kusaka, K. Simpson, A. G.  Williams, Phys. Rev. D56, 5071 (1997).
    
\bibitem{SAAD2013}
V. Sauli, J.  Adam Jr., Phys. Rev. D67, 085007 (2013).
 
\bibitem{SA2008}
V.  Sauli, J. Phys. G35, 035005 (2008).  
 
 \bibitem{KACA2006}   
V. A.  Karmanov, J.  Carbonell, Eur. Phys. J. A27, 1 (2006).
    
\bibitem{KACA2010}    
J. Carbonell, V.A. Karmanov, Eur. Phys. J. A46, 387 (2010).    
 
\bibitem{SAFAFR2017}
G. Salmè, W. de Paula, T. Frederico, M. Viviani, Few Body Syst. 58, no. 3, 118 (2017). 

\bibitem{AC2019}  
J. H. Alvarenga Nogueira, D. Colasante, V. Gherardi, T. Frederico, E. Pace, G. Salmè,  Phys. Rev. D 100, 016021 (2019).
     	
\bibitem{CAKA2011}    	
J. Carbonell, V.A. Karmanov, Few Body Syst. 49, 205-222,  (2011).

\bibitem{CO1982}
 J.M. Cornwall, Phys. Rev. D 26,1453 (1982).
  
\bibitem{VHP}
V. Sauli,   Hadron Vacuum Polarization from application of DSEs and analytical confinement, submitted for publication, ArXiv1809.07644.
 
\bibitem{PTF}
 V. Sauli, Gauge Technique approximation to the πγ production and the pion transition form factor, submitted for publication, ArXiv:1905.07221.
 
\bibitem{weinberg}
S. Weinberg, The Quantum Theory of Fields I, Cambriddge University Press.

\bibitem{Cabibbo}
N. Cabibbo and R. Gatto, Phys. Rev.224 , N.5.1577-1595 (1961).

\bibitem{salam63}
 A. Salam, Phys. Rev. 130, 1287 (1963)
 
\bibitem{stra64}
  J. Strathdee, Phys. Rev.135, 1428 (1964) 
  
\bibitem{DEWE1977}
  R. Delbourgo, P. C. West, J. Phys. A 10, 1049 (1977)
   
\bibitem{DEWE2}
 R. Delbourgo, P.C. West, Phys. Lett.B 72, 96 (1977)
   
\bibitem{DE1979}
   R. Delbourgo, Nuovo Cim.49, 484 (1979)

\bibitem{BISA2007}
  V. Sauli, J. Adam, P. Bicudo  Phys. Rev. D75. 087701 (2007). 
  






  
\end{thebibliography}
\end{document}